\documentstyle[prl,aps,epsfig]{revtex}         
\begin{document}
\draft               
\twocolumn[\hsize\textwidth\columnwidth\hsize\csname @twocolumnfalse\endcsname

\title{Spatial Semiconductor-Resonator Solitons}

\author{V. B. Taranenko, C. O. Weiss}

\address{Physikalisch-Technische Bundesanstalt \\
38116 Braunschweig, Germany}

\maketitle

\begin{abstract}
We demonstrate experimentally and numerically the existence of
spatial solitons in multiple-quantum-well semiconductor
microresonators driven by an external coherent optical field. We
discuss stability of the semiconductor-resonator solitons over a
wide spectral range around the band edge. We demonstrate the
manipulation of such solitons: switching solitons on and off by
coherent as well as incoherent light; reducing the light power
necessary to sustain and switch a soliton, by optical pumping.
\end{abstract}
 \vskip1pc ]
\section{Introduction}
Spatial resonator solitons theoretically predicted in
\cite{tag:1,tag:2,tag:3,tag:4,tag:5} can exist in a variety of
nonlinear resonators, such as lasers (vortices), lasers with
saturable absorber (bright solitons), parametric oscillators
(phase solitons), driven nonlinear resonators (bright/dark
solitons). Such resonator solitons can be viewed as self-trapped
domains of one field state surrounded by another state of the
field. The two different field states can be a high and a low
field (bright/dark solitons), positive and negative field (phase
solitons), right-hand and left-hand polarized field (polarization
solitons) or high and zero field (vortices).

Bright solitons in laser resonators with saturable absorber were
initially shown to exist in \cite{tag:6}, which was limited to
single stationary solitons. Existence of moving solitons and
simultaneous existence of large numbers of stationary solitons
was shown in \cite{tag:7,tag:8}. Phase solitons in degenerate
parametric wave mixing resonators were predicted in \cite{tag:9}
and demonstrated in \cite{tag:10}. Theoretically it was shown in
\cite{tag:11} that not only 2D spatial resonator solitons exist
but that also in 3D such structures can be stable, linking the
field of optical solitons with elementary particle physics
\cite{tag:12}. Vortex solitons differ from other soliton types in
that they possess \emph{structural stability} in addition to
\emph{dynamical stability}, the only stabilizing mechanism of the
other solitons. The existence of vortices in lasers initially
shown in \cite{tag:13} and later also the existence of vortex
solitons \cite{tag:14}.

Since the resonator solitons are bistable and can be moved around
they are suited to carry information. Information can be written
in the form of a spatial soliton, somewhere, and then transported
around at will; finally being read out somewhere else; possibly
in conjunction with other solitons. In this respect the spatial
resonator solitons have no counterpart in any other kind of
information-carrying elements and lend themselves therefore to
operations not feasible with conventional electronic means, such
as an all-optical pipeline storage register ("photon buffer") or
even processing in the form of cellular automata. Experiments on
the manipulation of bright solitons as required for such
processing tasks were first carried out on a slow system: laser
with (slow) saturable absorber. In particular it was demonstrated
how to write and erase solitons and how to move them or localize
them. For reviews see \cite{tag:15,tag:16}.

\begin{figure}[htbf] \epsfxsize=65mm
\centerline{\epsfbox{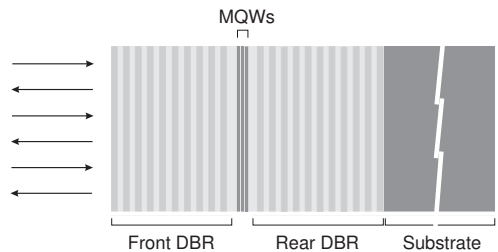}} \vspace{0.7cm}
\caption{Schematic of semiconductor microresonator consisting of
two plane distributed Bragg reflectors (DBR) and multiple quantum
wells (MQW).}
\end{figure}

In order to be applicable to technical tasks it is mandatory to
operate in fast, miniaturized systems. For compatibility and
integrability with other information processing equipment it is
desirable to use semiconductor systems. We chose the
semiconductor microresonator structure (Fig.~1) as commonly used
for vertical cavity surface emitting lasers (VCSEL) \cite{tag:17}
consisting of multiple quantum well (MQW) structure sandwiched
between distributed Bragg reflectors (DBR).

The resonator length of such a structure is $\sim$ $\lambda$
while the transverse size is typically 5 cm. Therefore such short
length and wide area microresonator permits only one longitudinal
mode (Fig.~2~(a)) and an enormous number of transverse modes that
allow a very large number of spatial solitons to coexist. The
resonator is obviously of the plane mirror type, implying
frequency degeneracy of all transverse modes and thus allowing
arbitrary field patterns to be resonant inside the resonator.
This is another prerequirement for existence and manipulability
of spatial solitons.

The resonator soliton existence is closely linked with the plane
wave resonator bistability (Fig.~2~(b)) caused by longitudinal
nonlinear effects: the nonlinear changes of the resonator length
(due to nonlinear refraction changes) and finesse (due to
nonlinear absorption changes) \cite{tag:18}. The longitudinal
nonlinear effects combined with transverse nonlinear effects
(such as self-focusing) can balance diffraction and form resonator
soliton. Generally these nonlinear effects can cooperate or act
oppositely, with the consequence of reduced soliton stability in
the later case.

\begin{figure}[htbf]
\epsfxsize=85mm \centerline{\epsfbox{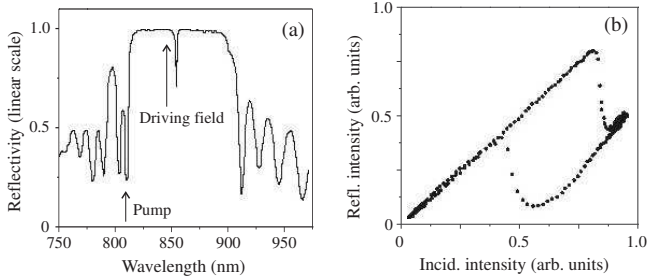}} \vspace{0.7cm}
\caption{Semiconductor microresonator reflectance spectrum (a)
and typical bistability loop in reflection (b). Arrows mark the
driving field that is detuned from the resonator resonance and
the pump field that is tuned to be coupled into resonator through
one of short-wavelength interference notches of the resonator
reflectance spectrum.}
\end{figure}

In the present paper we demonstrate experimentally and
numerically existence of bright and dark spatial solitons as well
as extended hexagonal patterns  in MQW-semiconductor
microresonators at room temperature. We discuss stability of the
semiconductor-resonator solitons over a wide spectral range
around the band edge. We demonstrate the manipulation of such
solitons in view of technical application: switching solitons on
and off by coherent as well as incoherent light; reducing the
light power necessary to sustain and switch a soliton, by optical
pumping.

\section{Model and numerical analysis}
We consider phenomenological model of a driven wide area
MQW-semiconductor microresonator similar to \cite{tag:19,tag:20}.
The optical field $E$ inside the resonator is described in
mean-field approximation \cite{tag:21}. The driving incident field
{$E_{\rm in}$} assumed to be a stationary plane wave. Nonlinear
absorption and refractive index changes induced by the
intracavity field in the vicinity of the MQW-structure band edge
are assumed to be proportional to the carrier density $N$
(normalized to the saturation carrier density). The equation of
motion for $N$ involves nonresonant (to avoid saturation effect)
pumping $P$, carrier recombination and diffusion. The resulting
coupled equations describing the spatio-temporal dynamics of $E$
and $N$ have the form:
\begin{eqnarray}
\cr {\partial E}/{\partial t}=E_{\rm in}-\sqrt{T}E\{[1+C{\rm
Im}(\alpha)(1-N)]+\cr{+i(\theta-C{\rm
Re}(\alpha)N-\nabla^{2}_{\bot})\}}\,, \cr\cr{\partial N}/{\partial
t}=P-\gamma[N-|E|^2(1-N)-d\nabla^{2}_{\bot}N]\,\,,
\end{eqnarray}
where $C$ is the saturable absorption scaled to the resonator DBR
transmission $T$ ($T$ is assumed to be small since the DBR
reflectivity is typically $\geq$ 0.995).  ${\rm Im}(\alpha)(1-N)$
and ${\rm Re}(\alpha)N$  describe the absorptive and refractive
nonlinearities, respectively, $\theta$  is the detuning of the
driving field from the resonator resonance, $\gamma$ is the photon
lifetime in the resonator scaled to the carrier recombination
time, $d$ is the diffusion coefficient scaled to the diffraction
coefficient and $\nabla^{2}_{\bot}$ is the transverse Laplacian.

Linear effects in the resonator are spreading of light by
diffraction and diffusion (terms with $\nabla^{2}_{\bot}$ in
(1)). The material nonlinearity that can balance this linear
spreading can do this in various ways. It has a real (refractive)
and imaginary (dissipative) part and can act longitudinally and
transversely. The nonlinear changes of the resonator finesse (due
to nonlinear absorption change) and length (due to nonlinear
refractive index change) constitute longitudinal nonlinear
effects, also known under the name \emph{nonlinear resonance}
\cite{tag:22}. The transverse effect of the nonlinear refractive
index can be self-focusing (favorable for bright and unfavorable
for dark solitons) and self-defocusing (favorable for dark and
unfavorable for bright solitons). Absorption (or gain) saturation
(bleaching) leading to nonlinear gain guiding in laser parlance,
a transverse effect. Longitudinal and transverse effects can work
oppositely, or cooperate.

There are two main external control parameters: the driving field
intensity $|E|^2$  and the resonator detuning $\theta$. Then for
driving intensities not quite sufficient for reaching the
resonance condition for the whole resonator area, the system
"chooses" to distribute the light intensity in the resonator in
isolated spots where the intensity is then high/low enough to
reach the resonance condition thus forming bright/dark patterns.
Instead of saying "the system chooses" one would more
mathematically express this by describing it as a modulational
instability (MI). The detuned plane wave field without spatial
structure with intensity insufficient to reach the resonance
condition is unstable against structured solutions. According to
our numerical solutions of (1) a large number of such structured
solutions coexist and are stable (see e.g. patterns in Fig.~7).

Fig.~3 shows the existence domains (in coordinates $\theta$ -
$|E|^2$) of MI, dark spatial solitons and plane-wave bistability
calculated for the case of purely dispersive (defocusing)
nonlinearity. The dark soliton structure (inset in Fig.~3) can be
interpreted as a small circular switching front. A switching
front connects two stable states: the high transmission and the
low transmission state. Such a front can in 2D surround a domain
of one state. When this domain is comparable in diameter to the
"thickness" of the front, then each piece of the front interacts
with the piece on the opposite side of the circular small domain,
which can lead, particularly if the system is not far from a
modulational instability (see Fig.~3), to a stabilisation of the
diameter of the small domain. In which case the small domain
becomes an isolated self-trapped structure or a dissipative
resonator soliton.

\begin{figure}[htbf]
\epsfxsize=60mm \centerline{\epsfbox{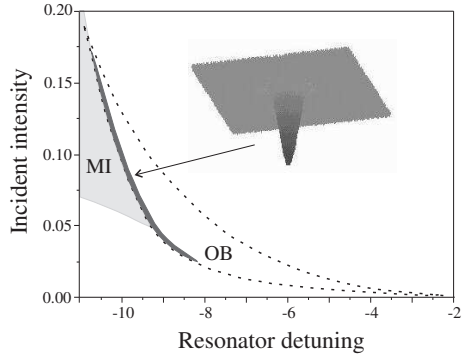}} \vspace{0.9cm}
\caption{Numerical solutions of (1) for unpumped ($P=0$) and
dispersive/defocusing (${\rm Im}(\alpha)=0$) case. Area limited
by dashed lines is optical bistability domain for plane waves.
Grey shaded area is modulational instability domain. Dark shaded
area is domain of stability for dark solitons. Inset is dark
soliton in 3D representation. Parameters: $C=10$, $T=0.005$,
${\rm Re}(\alpha)=-1$, $d=0.01$.}
\end{figure}

\section{Experimental technique}
Fig.~4 shows the optical arrangement for the
semiconductor-resonator soliton experiments, and in particular for
their switching on or off. The semiconductor microresonators
consisting of MQW (GaAs/AlGaAs or GaInAs/GaPAs) structures
sandwiched between high-reflectivity ($\geq$ 0.995) DBRs (Fig.~1)
operate at room temperature. The microresonator structures were
grown on the GaAs substrates by molecular beam epitaxy technique
that makes possible high quality MQW structures with small radial
layer thickness variation. The best sample tested in our
experiments has only $\sim$ 0.3 nm/mm variation of the resonator
resonance wavelength over the sample cross section.

The driving light beam was generated by either a tunable (in the
range 750-950 nm) Ti:Sa laser or a single-mode laser diode, both
operating in continuous wave regime. For experimental convenience
to limit thermal effects we perform the whole experiments within
a few microseconds, by admitting the light through an
acousto-optical modulator. The laser beam of suitable wavelength
is focused on the microresonator surface in the light spot of
$\sim$ 50 $\mu$m in diameter, thus providing a quite large Fresnel
number ($\geq$ 100).

\begin{figure}[htbf]
\epsfxsize=85mm \centerline{\epsfbox{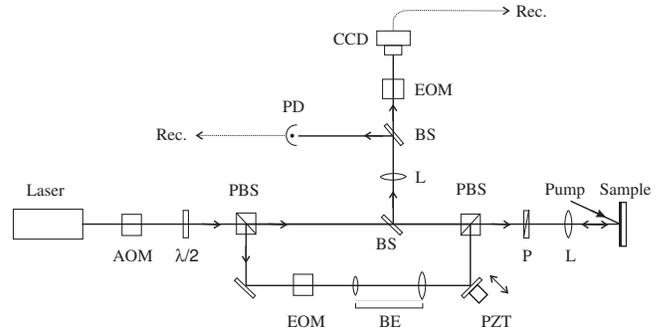}} \vspace{0.9cm}
\caption{Experimental setup. Laser: Ti:Sa (or diode) laser, AOM:
acousto-optic modulator, $\lambda$/2: halfwave plate, PBS:
polarization beam spliters, EOM: electro-optical amplitude
modulators, BE: beam expander, PZT: piezo-electric transducer, P:
polarizer, L: lenses, BS: beam splitters, PD: photodiode. }
\end{figure}

Part of laser light is split away from the driving beam and then
superimposed with the main beam by a Mach-Zehnder interferometer
arrangement, to serve as the address beam. The address beam is
sharply focused and directed to some particular location in the
illuminated. The switching light is opened only for a few
nanoseconds using an electro-optic modulator. For the case of
incoherent switching the polarization of the address beam is
perpendicular to that of the main beam to avoid interference. For
the case of coherent switching the polarizations should be
parallel and a phase control of the switching field is thus
always needed: for switching on as well as switching off a
soliton. One of the interferometer mirrors can be moved by a
piezo-electric element to control the phase difference between
the driving light and the address light.

Optical pumping of the MQW-structures was done by a multi-mode
laser diode or a single-mode Ti:Sa laser. To couple the pump
light into the microresonator the laser wavelength was tuned into
the high transmission spectral window of the microresonator
reflectance spectrum as shown in Fig.~2~(a).

The observation is done in reflection by a CCD camera combined
with a fast shutter (another electro-optic modulator), which
permits to take nanosecond snapshots at a given time of the
illuminated area on the resonator sample. Recording movies on
this nanosecond time scale is also possible. To follow intensity
in time in certain points (e.g. at the location of a soliton) a
fast photodiode can be imaged onto arbitrary locations within the
illuminated area.

\section{Experimental results and discussions}
To find the most stable resonator solitons for applications one
can play with the nonlinear (absorptive/dispersive) response by
choice of the driving field wavelength, with the resonator
detuning, and finally with the carrier population inversion by
the pumping. We recall that all nonlinearities change their sign
at transparency i.e. at the point where in the valence- and the
conduction band populations are equal. Going from below
transparency (absorption) to above transparency (population
inversion, producing light amplification), nonlinear absorption
changes to nonlinear gain, self-focusing changes to
self-defocusing and vice versa, and decrease of resonator length
with intensity changes to increase (and vice versa). The
population of the bands can be controlled by pumping ($P$ in (1))
i.e. transferring electrons from the valence band to the
conduction band. This can be done by optical excitation
\cite{tag:23}, with radiation of wavelength shorter than the band
edge wavelength or - if the structure is suited to support
electrical currents (i.e. if it is a real VCSEL-structure) - by
electrical excitation.

\subsection{Below bandgap hexagons and dark solitons}
Working well bellow the bandgap when the driving field wavelength
is $\sim$ 30 nm longer than the band edge wavelength we observed
spontaneous formation of hexagonal patterns (Fig.~5). The hexagon
period scales linearly with $\theta^{-1/2}$ \cite{tag:24}
indicating that they are formed by the tilted-wave mechanism
\cite{tag:25} that is the basic mechanism for resonator hexagon
formation \cite{tag:26}. Dark-spot hexagon (Fig.~5~(a)) converts
in to bright-spot hexagon (Fig.~5~(b)) when the driving intensity
increases. Experiment shows that individual spots of these
patterns can not be switched on/off independently from other
spots as it is expected for strongly correlated spot structure
(or coherent hexagons).

\begin{figure}[htbf]
\epsfxsize=50mm \centerline{\epsfbox{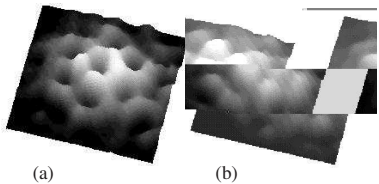}} \vspace{0.7cm}
\caption{(a) Bright (dark in reflection) and (b) dark (bright in
reflection) hexagonal patterns for the dispersive/defocusing
case. Driving light intensity incrieses from (a) to (b). }
\end{figure}

However when we further increased the driving intensity we found
that the bright spots in such hexagonal patterns can be switched
independently by the addressed focused optical (incoherent) pulses
\cite{tag:24}. Fig.~6 shows the experimental results. Fig.~6~(a)
shows the hexagonal pattern formed. The focused light pulse can
be aimed at individual bright spots such as the ones marked "1"
or "2". Fig.~6~(b) shows that after the switching pulse aimed at
"1" spot "1" is off. Fig.~6~(c) shows the same for spot "2". We
remark that in these experiments we speak of true logic
switching: the spots remain switched off after the switching
pulse (if the energy of the pulse is sufficient, otherwise the
bright spot reappears after the switching pulse). These
observations of local switching indicate that these hexagonal
patterns are not coherent pattern: the individual spots are
rather independent, even at this dense packing where the spot
distance is about the spot size.

\begin{figure}[htbf]
\epsfxsize=70mm \centerline{\epsfbox{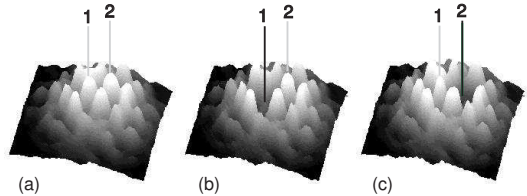}} \vspace{0.7cm}
\caption{Switching-off of individual spots of incoherent
hexagonal structure with addressed pulses located at different
places (marked 1 and 2) of the pattern. }
\end{figure}

These experimental findings can be understood in the frame of the
model (1) \cite{tag:27}. Fig.~7 shows the bistable plane wave
characteristic of the semiconductor resonator for conditions
roughly corresponding to the experimental conditions. At the
intensities marked (a) to (d) patterned solutions exist.

\begin{figure}[htbf]
\epsfxsize=60mm \centerline{\epsfbox{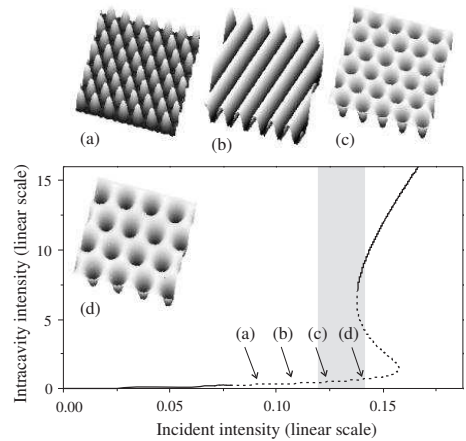}} \vspace{0.7cm}
\caption{Numerical solution of (1) for intracavity light
intensity as function of incident intensity: homogeneous solution
(dashed line marks modulationaly unstable part of the curve) and
patterns (a-d). Shaded area marks existence range for dark-spot
hexagons. Parameters: $C=10$, $T=0.005$, ${\rm Im}(\alpha)=0.1$,
${\rm Re}(\alpha)=-1$, $\theta=-10.3$, $d=0.01$. }
\end{figure}

The pattern period in Fig.~7~(a) corresponds precisely to the
detuning in the following way. When the driving field is detuned,
the resonance condition of the resonator cannot be fulfilled by
plane waves travelling exactly perpendicularly to the mirror
plane. However, the resonance condition can be fulfilled if the
wave plane is somewhat inclined with respect to the mirror plane
(the tilted wave solution \cite{tag:25}). The nonlinear system
chooses therefore to support resonant, tilted waves. Fig.~7~(a) is
precisely the superposition of 6 tilted waves that support each
other by (nonlinear) 4-wave-mixing. The pattern period
corresponds to the resonator detuning as in the experiment for
structures Fig.~5~(a). Thus the pattern formation in Fig.~7~(a) is
mostly a linear process. In this pattern the bright spots are not
independent. Individual spots cannot be switched as in the
experiment Fig.~5.

On the high intensity pattern Fig.~7~(d) the pattern period is
remarkably different from Fig.~7~(a) even though the detuning is
the same. This is indication that the internal detuning is
smaller and means that the resonator length is nonlinearly
changed by the intensity-dependent refractive index (the nonlinear
resonance). From the ratio of the pattern periods of Fig.~7~(a)
and (d) one sees that the nonlinear change of detuning is about
half of the external detuning. That means the nonlinear detuning
is by no means a small effect. This in turn indicates that by
spatial variation of the resonator field intensity the detuning
can vary substantially in the resonator cross section. In other
words, the resonator has at the higher intensity a rather wide
freedom to (self-consistently) arrange its field structure. One
can expect that this would allow a large number of possible
stable patterns between which the system can choose.

\begin{figure}[htbf]
\epsfxsize=70mm \centerline{\epsfbox{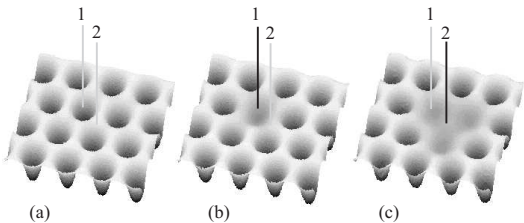}} \vspace{0.7cm}
\caption{Stable hexagonal arrangements of dark patial solitons:
(a) without defects, (b) with single-soliton defect, (c) with
triple-soliton defect.}
\end{figure}

Fig.~8 shows that the high intensity conditions of Fig.~7~(d)
allow reproducing the experimental findings on switching
individual bright spots. Fig.~8~(a) is the regular hexagonal
pattern, Fig.~8~(b) shows one bright spot switched off as a
stable solution and Fig.~8~(c) shows a triple of bright spots
switched off as a stable solution, just as observed in the
experiments \cite{tag:27}.

Thus while Fig.~7~(a) is a completely coherent space filling
pattern, Fig.~7~(d) is really a cluster of (densest packed)
individual dark solitons. The increase of intensity from (a) to
(d) allows the transition from the extended patterns to the
localized structures by the increased nonlinearity, which gives
the system an additional internal degree of freedom. We note that
the transition from the coherent low intensity pattern to the
incoherent higher intensity structure proceeds through
stripe-patterns as shown in Fig.~7~(b) \cite{tag:27}. For the
intensity of Fig.~7~(c) the individual spots are still not
independent as in the experiment Fig.~5~(b).

\subsection{Near bandgap bright and dark solitons}
Working at wavelengths close to the band edge we found both
bright and dark solitons, as well as collections of several of
the bright and dark solitons, several solitons existing at the
same time (Figs~9,~10). Fig.~10 demonstrates that shape and size
of bright spots are independent on the shape and intensity of the
driving beam that is a distinctive feature of spatial solitons.

\begin{figure}[htbf]
\epsfxsize=65mm \centerline{\epsfbox{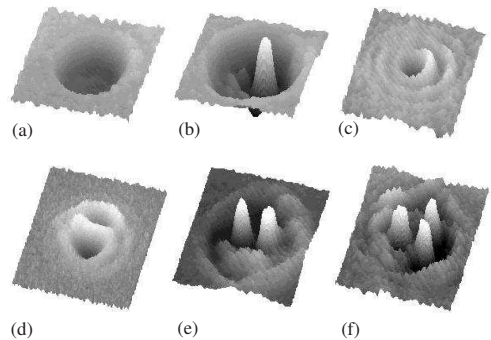}} \vspace{0.7cm}
\caption{Solitons in the semiconductor microresonator: (a)
switched area without solitons (for completeness); (b) dark
soliton in switched area; (c) bright soliton on unswitched
background; (d) 2 bright solitons; (e), (f) 2,3 dark solitons.}
\end{figure}

\begin{figure}[htbf]
\epsfxsize=75mm \centerline{\epsfbox{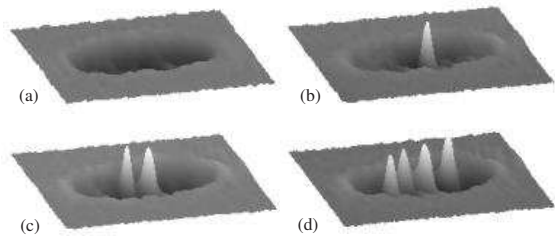}} \vspace{0.7cm}
\caption{Switched domain (a) and dark solitons (b)-(d) in an
illuminating area of elliptical shape.}
\end{figure}

Nonlinearity of the MQW structure near the band edge is
predominantly absorptive. Therefore in the first approximation we
can neglect the refractive part of the complex nonlinearity in
the model equations (1) and describe the nonlinear structure as a
saturable absorber. Numerical simulations for this case (Fig.~11)
confirm existence of both bright and dark resonator solitons as
they are observed in the experiment (Figs~9,~10). We can contrast
these resonator spatial solitons with the propagating (in a bulk
nonlinear material) spatial solitons: the latter can not be
supported by a saturable absorber.

\begin{figure}[htbf]
\epsfxsize=60mm \centerline{\epsfbox{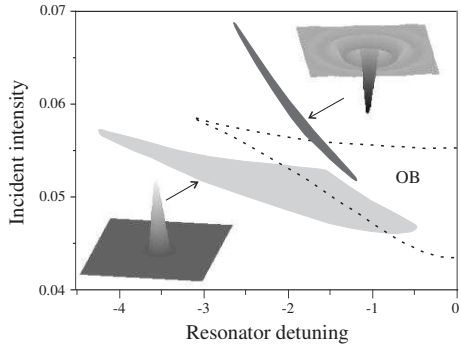}} \vspace{0.7cm}
\caption{Numerical solutions of (1) for unpumped ($P=0$) and
absorptive (${\rm Re}(\alpha)=0$) case. Area limited by dashed
lines is optical bistability domain for plane waves. Shaded areas
are domains of existence of bright (grey area) and dark (black
area) solitons. Insets are bright (left) and dark (right)
soliton. Parameters: $C=20$, $T=0.005$, ${\rm Im}(\alpha)=1$,
$d=0.01$.}
\end{figure}

Fig.~12~(c) and (d) show details of the spontaneous formation of
the bright soliton (as in Fig.~9~(c)). As discussed in
\cite{tag:28} temperature effects lead in this case to a slow
formation of solitons, associated with the shift of the band edge
by temperature \cite{tag:29}. Fig.~12~(c) gives the incident and
reflected intensity at the location of the bright soliton and
Fig.~12~(d) gives the reflectivity on a diameter of the
illuminated area. At 1.3 $\mu$s (arrow) the resonator switches to
high transmission (low reflection). The switched area then
contracts relatively slowly to the stable structure Fig.~12~(d),
which is existing after $t$ $\approx$ 3.0 $\mu$s.

After the resonator switches to low reflection its internal field
and with it the dissipation is high. A rising temperature
decreases the band gap energy \cite{tag:29} and therefore shifts
the bistable resonator characteristic towards higher intensity.
Thus the basin of attraction for solitons which is located near
the plane wave switch-off intensity (see locations of the
existence domains for the bright solitons and the plane wave
bistability in Fig.~11) is shifted to the incident intensity,
whereupon a soliton can form. Evidently for different parameters
the shift can be substantially larger or smaller than the width of
bistability loop, in which case no stable soliton can appear. We
note that in absence of thermal effects (good heat-sinking of
sample) solitons would not appear spontaneously but would have to
be switched on by local pulsed light injection.

\begin{figure}[htbf]
\epsfxsize=70mm \centerline{\epsfbox{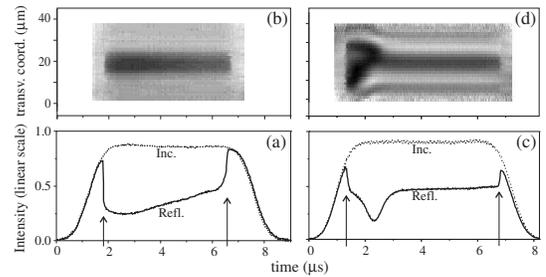}} \vspace{0.7cm}
\caption{Comparison of bright soliton formation above bandgap
(left) and below bandgap (right). Reflectivity on a diameter of
the illuminated area as a function of time (b), (d). Intensity of
incident (dotted) and reflected (solid) light, at the center of
the soliton as a function of time (a), (c). Arrows mark the
switch-on and -off.}
\end{figure}

Fig.~13~(a) shows the incoherent switch-on of bright soliton,
where the perpendicularly polarized switching pulse ($\sim$ 10 ns)
is applied at $t$ = 4 $\mu$s. As apparent, a soliton forms after
this incoherent light pulse. The slow formation of the soliton is
apparent in Fig.~13~(a) (using roughly the time from $t$ = 4 to
$t$ = 4.5 $\mu$s).

It should be emphasized that this thermal effect is not
instrumental for switching a soliton on. However, it allows
switching a soliton off incoherently \cite{tag:30}. This is shown
in Fig.~13~(b) where the driving light is initially raised to a
level at which a soliton forms spontaneously (note again the slow
soliton formation due to the thermal effect). The incoherent
switching pulse is then applied which leads to disappearance of
the soliton.

\begin{figure}[htbf]
\epsfxsize=60mm \centerline{\epsfbox{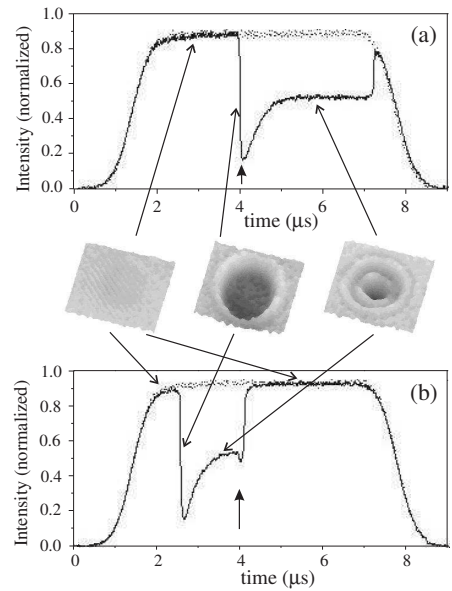}} \vspace{0.7cm}
\caption{Recording of incoherent switching-on (a) and
switching-off (b) of a soliton. Snapshot pictures show unswitched
state (left), circular switched domain (center). and a soliton
(right). Dotted trace: incident intensity.}
\end{figure}

The soliton can thus be switched on and also off by an incoherent
pulse. The reason for the latter is thermal. Initially the
material is "cold". A switching pulse leads then to the creation
of a soliton. Dissipation in the material raises the temperature
and the soliton is slowly formed. At the raised temperature the
band edge (and with it the bistability characteristic) and the
existence range of solitons is shifted so that a new pulse brings
the system out of the range of existence of solitons.
Consequently the soliton is switched off.

Thus switching on a soliton is possible incoherently with the
"cold" material and switching off with the "heated" material.
When the driving intensity is chosen to be slightly below the
spontaneous switching threshold the nonlinear resonator is cold.
An incoherent pulse increases illumination locally and can switch
the soliton on that causes local heating of the resonator. Another
incoherent pulse aimed into the heated area can then switch the
soliton off and thereby cool the resonator to its initial state,
so that the soliton could be switched on/off again.

\subsection{Above bandgap bright solitons}
At excitation above bandgap bright solitons form analogously to
the below/near bandgap case \cite{tag:31}. Fig.~14 shows the
bright solitons as observed in the reflected light with their
characteristic concentric rings with the same appearance as the
bright solitons below band gap (Fig.~9~(c)).

\begin{figure}[htbf]
\epsfxsize=50mm \centerline{\epsfbox{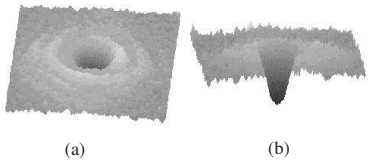}} \vspace{0.7cm}
\caption{Bright soliton above bandgap in 3D representation: view
from above (a), from below (b).}
\end{figure}

Fig.~12 compares the dynamics of the bright soliton formation for
excitation above bandgap (left) and below bandgap (right).
Difference between these two cases can be understood from the
model. From (1) we obtain the reflected intensity as a function of
incident intensity for wavelengths above (${\rm Re}(\alpha)>0$),
as well as below the bandgap (${\rm Re}(\alpha)<0$), for plane
waves (Fig.~15). One sees that the bistability range is large
below and small above the bandgap. Solving (1) numerically the
typical bright soliton (top of Fig.~15) is found coexisting with
homogeneous intensity solutions in the shaded regions of
Fig.~15~(a), (b).

After switching on the resonator below band gap (Fig.~15~(b)), the
intensity in the resonator is high and with it the thermal
dissipation. The temperature consequently rises, which shifts the
band gap \cite{tag:29} and with it the bistability characteristic,
so that the switch-off intensity, close to which the stable
solitons exist, becomes close to the incident intensity. Then the
resonator is in the basin of attraction for the solitons and the
soliton forms as observed in Fig.~12~(c), (d).

\begin{figure}[htbf]
\epsfxsize=85mm \centerline{\epsfbox{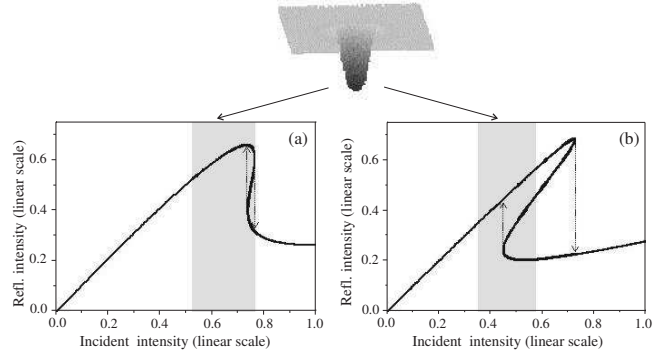}} \vspace{0.7cm}
\caption{Steady-state plane wave solution of (1) above bandgap
(a): ${\rm Re}(\alpha)=0.05$; and below band gap (b): ${\rm
Re}(\alpha)=-0.05$. Other parameters: $C=30$, ${\rm
Im}(\alpha)=0.99$, $\theta=-3$, $P=0$, $d=0.1$. The soliton
solution shown exists for incident intensities corresponding to
the shaded areas, in coexistence with homogeneous solutions. For a
temperature increase the characteristics together with soliton
existence ranges shift to higher incident intensities. Reflected
and incident intensities normalized to the same value.}
\end{figure}

\begin{figure}[htbf]
\epsfxsize=80mm \centerline{\epsfbox{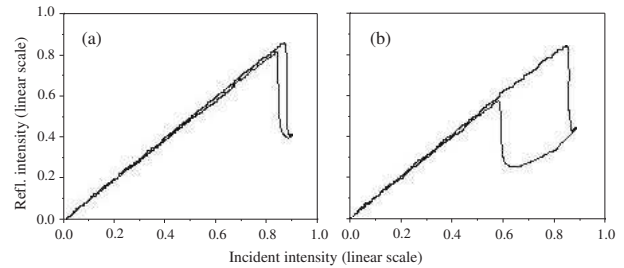}} \vspace{0.7cm}
\caption{Bistability characteristics (reflected light intensity
versus incident light intensity) measured at the center of an
illuminated area of 100 $\mu$m diameter, for above band gap (a)
and below band gap (b) excitation.}
\end{figure}

Above the band gap Fig.~12~(a), (b) show that the soliton is
switched on "immediately" without the slow thermal process. Fig.~
15~(a) shows why. The plane wave characteristic of the resonator
above band gap is either bistable but very narrow, or even
monostable (due to the contribution of the self-focusing reactive
nonlinearity \cite{tag:32}) but still with bistability between the
soliton state (not plane wave) and the unswitched state. In this
case the electronic switching leads directly into the basis of
attraction for solitons and the switch-on of the soliton is
purely electronic and fast. The width of the bistability
characteristics observed experimentally (Fig.~16) scale in
agreement with Fig.~15.

Nonetheless, also above bandgap there is strong dissipation after
the switch-on. The associated temperature rise influences and can
even destabilize the soliton. The effect can be seen in
Fig.~12~(a). Over a time of a few $\mu$s after the soliton
switch-on the soliton weakens (reflectivity increases slowly)
presumably by the rise of temperature and the associated shift of
the band gap. At 6.5 $\mu$s the soliton switches off although the
illumination has not yet dropped.

Thus, while the dissipation does not hinder the fast switch-on of
the soliton, it finally destabilizes the soliton. After the
soliton is switched off, the material cools and the band gap
shifts back so that the soliton could switch on again.

\begin{figure}[htbf]
\epsfxsize=60mm \centerline{\epsfbox{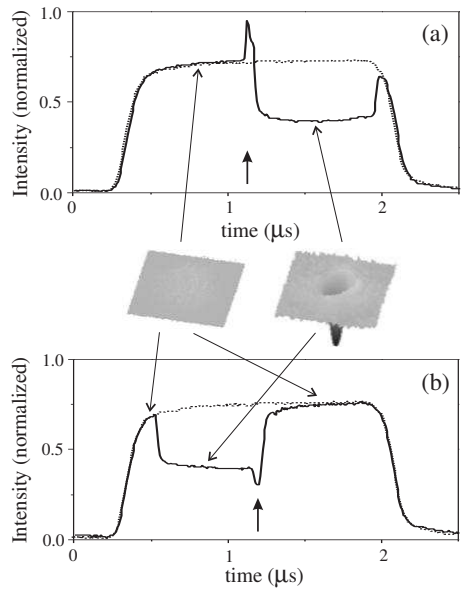}} \vspace{0.7cm}
\caption{Recording of coherent switching-on (a) and switching-off
(b) of a bright soliton. Heavy arrows mark the application of
switching pulses. Dotted traces: incident intensity. The insets
show intensity snapshots, namely unswitched state (left) and
soliton (right).}
\end{figure}

Fig.~17 shows the soliton coherent switching observations
\cite{tag:33}. Fig.~17~(a) shows switching a soliton on. The
driving light intensity is chosen slightly below the spontaneous
switching threshold. At t $\approx$ 1.2 $\mu$s the addressing
pulse is applied. It is in phase with the driving light, as
visible from the constructive interference. A bright soliton
results, showing up in the intensity time trace as a strong
reduction of the reflected intensity. Fig.~17~(b) shows switching
a soliton off. The driving light is increased to a level where a
soliton is formed spontaneously. The addressing pulse is then
applied in counterphase to the driving light, as visible from the
destructive interference. The soliton then disappears, showing up
in the intensity time trace as reversion of the reflected
intensity to the incident intensity value. The Fig.~17 insets
show 2D snapshots before and after the switching pulses for
clarity.

\subsection{Optical pumping}
The thermal effects discussed above result from the local heating
caused by the high intracavity intensity at the bright soliton
location. They limit the switching speed of solitons and they
will also limit the speed at which solitons could be moved
around, limiting applications. The picture is that a soliton
carries with it a temperature profile, so that the temperature
becomes a dynamic and spatial variable influencing the soliton
stability.

As opposed, a spatially uniform heating will not cause such
problems, as it shifts parameters but does not constitute a
variable in the system. The largely unwanted heating effects are
directly proportional to the light intensity sustaining a
soliton. For this reason and quite generally it is desirable to
reduce the light intensities required for sustaining solitons.

Conceptually this can be expected if part of the power sustaining
a resonator soliton could be provided incoherently to the driving
field, e.g. by means of optical pumping. Optical pumping of MQW
structure generates carriers and allows converting from
absorption to gain. In the last case the semiconductor
microresonator operates as VCSEL \cite{tag:17}. To couple pump
light into the microresonator the pump laser was tuned in one of
short-wavelength transparency windows in the reflection spectrum
of the microresonator as shown in Fig.~2~(a).

When the pumping was below the transparency point and the driving
laser wavelength was set near the semiconductor MQW structure
band edge the bright and dark solitons formed (Fig.~18~(a), (b))
\cite{tag:34} similar to the unpumped case Figs 9, 10.

\begin{figure}[htbf]
\epsfxsize=70mm \centerline{\epsfbox{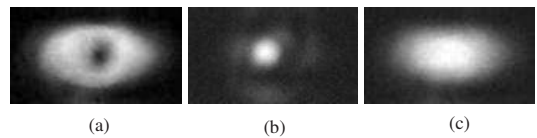}} \vspace{0.7cm}
\caption{Intensity snapshots of structures observed in reflection
from pumped (below transparency) semiconductor microresonator
illuminated near resonance showing bright (a) and dark (b)
soliton. The illuminating beam from the laser diode has an
elliptical shape (c).}
\end{figure}

Analysis of (1) shows that increase of the pump intensity leads to
shrinking of the resonator solitons' existence domain and
shifting towards low intensity of the light sustaining the
solitons. Such reduction of the sustaining light intensity was
observed experimentally in \cite{tag:23}. That is why soliton
switchings in the pumped case are fast \cite{tag:23} and not
mediated by thermal effects as for soliton formation without
pumping (Fig.~12~(c), Fig.~13).

When the pump intensity approaches the transparency point of the
semiconductor material, the resonator solitons' domain of
existence disappears. It reappears above the transparency point.
In the experiment we have quite strong contribution of the
imaginary part (absorption/gain) of the complex nonlinearity at
the working wavelength (near band edge). Therefore the
transparency point is very close to the lasing threshold so that
inversion without lasing is difficult to realize.

\begin{figure}[htbf]
\epsfxsize=70mm \centerline{\epsfbox{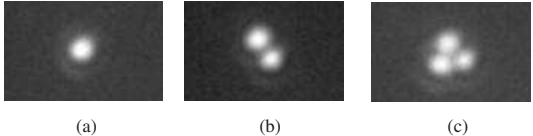}} \vspace{0.7cm}
\caption{Intensity snapshots of typical beam structures at
optical pump intensities slightly above lasing threshold (pump
increases from (a) to (c)).}
\end{figure}

Slightly above threshold we observe in presence of illumination
structures (Fig.~19) reminiscent of the solitons in electrically
pumped resonators \cite{tag:35}. We note that optical as opposed
to electrical pumping allows more homogeneous pumping conditions
\cite{tag:36}. This suggests that optically pumped resonators
lend themselves more readily for localization and motion control
of solitons then electrically pumped ones.

\begin{figure}[htbf]
\epsfxsize=70mm \centerline{\epsfbox{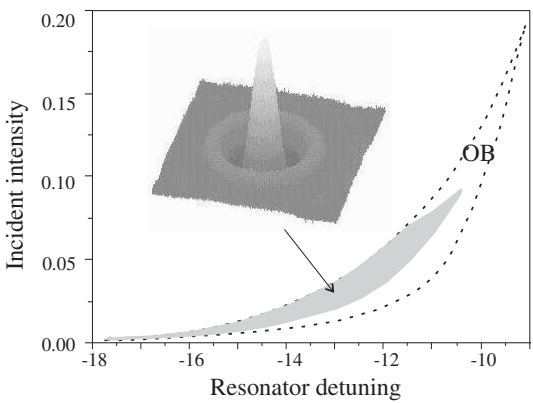}} \vspace{0.7cm}
\caption{Results of numerical simulations of below bandgap
(purely dispersive) solitons using the model (1) for pumped above
the transparency point microresonator ($P=2$). Inset is bright
soliton. Shaded area is domain of existence of resonator bright
solitons. Area limited by dashed lines is optical bistability
domain for plane waves. Other parameters same as in Fig.~3.}
\end{figure}

There is difference between below bandgap resonator solitons in
pumped and unpumped cases. The nonlinear resonance mechanism of
soliton formation \cite{tag:22} requires a defocusing
nonlinearity below transparency (Fig.~3) and a focusing
nonlinearity above transparency. Transverse nonlinear
self-focusing effect is generally furthering soliton formation.
Fig.~20 shows typical examples of calculated resonator solitons
for pumped semiconductor microresonator. Bright solitons have a
large existence range in the pumped case (Fig.~20), dark solitons
exist, though with smaller range of stability, in the unpumped
case (Fig.~3).

Thus optically pumped semiconductor resonators are well suited
for sustaining solitons below bandgap: (i) background light
intensity necessary to sustain and switch resonator solitons is
substantially reduced by the pumping and therefore destabilizing
thermal effects are minimized, (ii) above the transparency point
only the dispersive part of semiconductor nonlinearity stabilizes
a soliton, therefore the domain of existence of below bandgap
(purely dispersive) bright solitons can be quite large.

\section{Conclusion}
In conclusion, we have shown experimentally and numerically
existence of spatial solitons in driven semiconductor
microresonators over a wide spectral range around the bandgap:
below bandgap hexagons and dark solitons, near bandgap bright and
dark solitons, above bandgap bright solitons, optically pumped
below bandgap bright solitons. We have demonstrated the
manipulation of such solitons: switching them on and off by
coherent as well as incoherent light; reducing the light power
necessary to sustain and switch a soliton, by optical pumping.\\

Acknowledgment\\ This work was supported by Deutsche
Forschungsgemeinschaft under grant We743/12-1.


\begin{thebibliography}{99}
\vspace*{-0.7cm}
\bibitem{tag:1} D.W. Mc Laughlin, J.V. Moloney, A.C. Newell: Phys. Rev. Lett. \textbf{51}, 75 (1983)

\bibitem{tag:2} N.N. Rosanov, G.V. Khodova: Opt. Spectrosk. \textbf{65}, 1375 (1988)

\bibitem{tag:3} M. Tlidi, P. Mandel, R. Lefever: Phys. Rev. Lett. \textbf{73}, 640 (1994)

\bibitem{tag:4} G.S. Donald, W.J. Firth: J. Opt. Soc. Am. B \textbf{73}, 1328 (1990)

\bibitem{tag:5} M. Brambilla, L.A. Lugiato, M. Stefani: Europhys. Lett. \textbf{34}, 109 (1996)

\bibitem{tag:6} V.Yu Bazhenov, V.B. Taranenko, M.V. Vasnetsov: `Transverse optical effects in bistable active cavity with nonlinear absorber
on bacteriorhodopsin'. In: \emph{Proc. SPIE}, \textbf{1840},
(1992) pp. 183-193

\bibitem{tag:7} K. Staliunas, V.B. Taranenko, G. Slekys, R. Viselga, C.O. Weiss: Phys. Rev. A \textbf{57}, 599 (1998)

\bibitem{tag:8} G. Slekys, K. Staliunas, C.O. Weiss: Opt. Comm. \textbf{149}, 113 (1998)

\bibitem{tag:9} K. Staliunas, V.J. Sanchez-Morcillo: Phys. Rev. A \textbf{57}, 1454 (1998)

\bibitem{tag:10} V.B. Taranenko, K. Staliunas, C.O. Weiss: Phys. Rev. Lett. \textbf{81}, 2236 (1998)

\bibitem{tag:11} K. Staliunas: Phys. Rev. Lett. \textbf{81}, 81 (1998)

\bibitem{tag:12} C. Rebbi, G. Soliani: \emph{Solitons and Particles}, (World Scientific, 1984)

\bibitem{tag:13} C. Tamm: Phys. Phys. Rev. A \textbf{38}, 5960 (1988)

\bibitem{tag:14} K. Staliunas, C.O. Weiss, G. Slekys: `Optical vortices in lasers'. In: \emph{Optical Vortices}. ed.
by M. Vasnetsov and K. Staliunas (Nova Science Publishers, 1999)
pp. 125-182

\bibitem{tag:15} C.O. Weiss, M. Vaupel, K. Staliunas, G. Slekys, V.B. Taranenko: Appl. Phys. B \textbf{68}, 151 (1999)

\bibitem{tag:16} \emph{Soliton Driven Photonics}, ed. by A.D. Boardman, A.P. Sukhorukov (Kluver Academic Publishers, 2001)

\bibitem{tag:17} T.E. Sale: \emph{Vertical Cavity Surface Emitting Lasers}, (Wiley, 1995)

\bibitem{tag:18} H.M. Gibbs: \emph{Optical Bistability - Controlling Light with Light}, (Academic Press, 1985)

\bibitem{tag:19} L. Spinelli, G. Tissoni, M. Brambilla, F. Prati, L.A. Lugiato: Phys. Rev. A \textbf{58}, 2542 (1998)

\bibitem{tag:20} D. Michaelis, U. Peschel, F. Lederer: Phys. Rev. A \textbf{56}, R3366 (1997)

\bibitem{tag:21} L.A. Lugiato, R. Lefever: Phys. Rev. Lett. \textbf{58}, 2209 (1987)

\bibitem{tag:22} K. Staliunas, V.J. Sanchez-Morcillo: Opt. Comm. \textbf{139}, 306 (1997)

\bibitem{tag:23} V.B. Taranenko, C.O. Weiss, W. Stolz: Opt. Lett. \textbf{26}, 1574 (2001)

\bibitem{tag:24} V.B. Taranenko, I. Ganne, R. Kuszelewicz, C. O. Weiss: Phys. Rev. A \textbf{61}, 063818 (2000)

\bibitem{tag:25} P.K. Jacobsen, J.V. Moloney, A.C. Newell, R. Indik:  Phys. Rev. A \textbf{45}, 8129 (1992)

\bibitem{tag:26} W.J. Firth, A.J. Scroggie: Europhys. Lett. \textbf{26}, 521 (1994)

\bibitem{tag:27} V.B. Taranenko, C.O. Weiss, B. Sch\"apers: Phys. Rev. A \textbf{65}, 013812 (2002)

\bibitem{tag:28} V.B. Taranenko, I. Ganne, R. Kuszelewicz, C.O. Weiss: Appl. Phys. B \textbf{72}, 377 (2001)

\bibitem{tag:29} T. Rossler, R.A. Indik, G.K. Harkness, J.V. Moloney, C.Z. Ning: Phys. Rev. A \textbf{58}, 3279 (1998)

\bibitem{tag:30} V.B. Taranenko, C.O. Weiss: Appl. Phys. B \textbf{72}, 893 (2001)

\bibitem{tag:31} V.B. Taranenko, C.O. Weiss, W. Stolz:  J. Opt. Soc. Am. B \textbf{19}, 8129 (1992)

\bibitem{tag:32} S.H. Park, J.F. Morhange, A.D. Jeffery, R.A. Morgan, A. Chavez-Pirson, H.M. Gibbs, S.W. Koch, N. Peyghambarian, M.
Derstine, A.C. Gossard, J.H. English, W. Weidmann: Appl. Phys.
Lett. \textbf{52}, 1201 (1988)

\bibitem{tag:33} V.B. Taranenko, F.-J. Ahlers, K. Pierz: Appl. Phys. B
(2002) in print

\bibitem{tag:34} V.B. Taranenko, C.O. Weiss: `Spatial solitons in an optically pumped semiconductor microresonator'. nlin.PS/0204048 (2002)

\bibitem{tag:35} Report as given in www.pianos-int.org

\bibitem{tag:36} W.J. Alford, T.D. Raymond, A.A. Allerman:  J. Opt. Soc. Am. B \textbf{19}, 663 (1992)



\end{thebibliography}
\end{document}